# With no Color and Scent (part III):
# Architecture of metal "shells" grown on templates by pulse current electrodeposition

**Strukova G.K., Strukov G.V., Egorov S.V.**

Institute of Solid State Physics RAS, Academician Osipyan Str.2, 142432, Chernogolovka, Russia

**Abstract**
A method of growing mushroom- or shell-like nanostructured metal convex-concave models has been worked out. Silver, copper, nickel, rhodium and Pd-Ni, Pd-Co alloy structures are reproduced as a result of self-assembly of nanowires growing on porous membranes in the course of pulse current electrodeposition. It is shown that the method allows to model not only the shell shape but also the hierarchical structure at the nano-, micro- and meso-levels. A 1.2 mm-sized "shell" was grown from Pd-Ni alloy. The architecture of the models was studied by fragmentation and chemical etching. The images were obtained using SUPRA -50 VP and JEOL scanning electron microscopes. The metal shell is a bowl-shaped frame, its walls composed of densely packed nanoelements. Each nanoelement is a conical bundle of nanowires grown as a supported "wine glass". The shell inner surface is a weave of nanowires with a vegetation pattern with bottom-up directed lines. The inner surface exhibits also well-pronounced transverse rings formed by the bottom-up growing bundles of nanowires that compose the shell frame. A number of nanowire bundle ends rise to the shell outer surface as "nanoflowers" that can serve as templates for growing nanowires. In some cases the nanowires growing on the outer surface form a copy of plate mushroom. A hypothesis is proposed that pulsed growth on membranes is a tool of morphogenesis of many mushrooms and plants.

### 1. Introduction

In recent years methods of producing bio-inspired models have become an increasingly urgent research issue in nanoscience and nanotechnology [1]. Researchers' interest has been attracted by "nanoflowers", anisotropic semiconductor and metal nanostructures. For instance, hyperphotoluminescence on zinc oxide-based "nanoflowers" [2], the effect of giant Raman scattering on gold [3] and catalytic activity on platinum "nanoflowers" [4, 5] have been reported. A number of papers are devoted to studies of the structure of natural shells of molluscan shellfish exhibiting the mechanical properties that are so far unfeasible for synthesized materials and to development of methods of their model synthesis. [6, 7]. However, the known isolated cases of synthesis of metallic models of natural objects, including "nanoflowers", provide no information on general methods that enable synthesis and control of diverse structures. In our previous works it was shown that pulse current electrodeposition on porous membranes allows growing nano-meso-structures from metals and alloys which remarkably resemble natural objects: plants, mushrooms, shells [8, 9]. The feasibility of model shaping was also shown: in certain conditions mushroom- and shell-like convexo-concave models of the same kind were obtained [9].

This paper is devoted to a more detailed investigation of the architecture and hierarchic structure of metallic mushroom and shell models produced on templates by pulse current electrodeposition.

## 2. Experimental

The electrolytes and modes of electrodeposition of Pd-Ni and other metals were described in detail in [8, 9]. SEM investigations of the obtained samples were carried out on SUPRA 50VP and JEOL scanning electron microscope. The architecture and hierarchic structure of the "shells" were studied by comparing the images of the models with those of the pieces obtained by mechanical fragmentation or chemical dissolution (etching). Fragmentation was achieved by storage of alcohol "shells" suspension in an ultrasonic bath for 0.5-1.0 h. The "shell" surface layer was etched with acetic and nitric acids (70/30 vol.%) at 30° C for 30 s followed by water and alcohol washing.

## 3. Results and discussion
### 3.1. Architecture and hierarchic structure of metallic "shells"

Electrodeposition of Pd-Ni alloy on porous aluminum oxide membranes with disordered pore arrangement gave rise to diverse "vegetable" structures ("cauliflower", "broccoli", "squash" [8]), ensembles of densely packed nanowires in the form of agaric or fungus on trees [10] as well as elegant convex-concave shall-, lotus leaf -, cabbage leafs - and mushroom-like structures. We have managed to reproduce the structures using specially prepared polymer membranes in combination with the found electrodeposition modes (Fig.1).

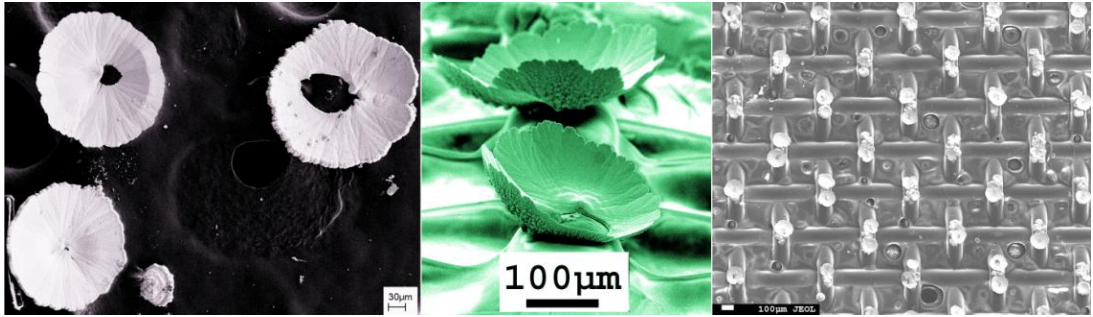

Fig.1. "Shells" and "lotus leaves".

Fig. 2 shows several typical convex-concave Pd-Ni (a-d) and Ag (f) models.

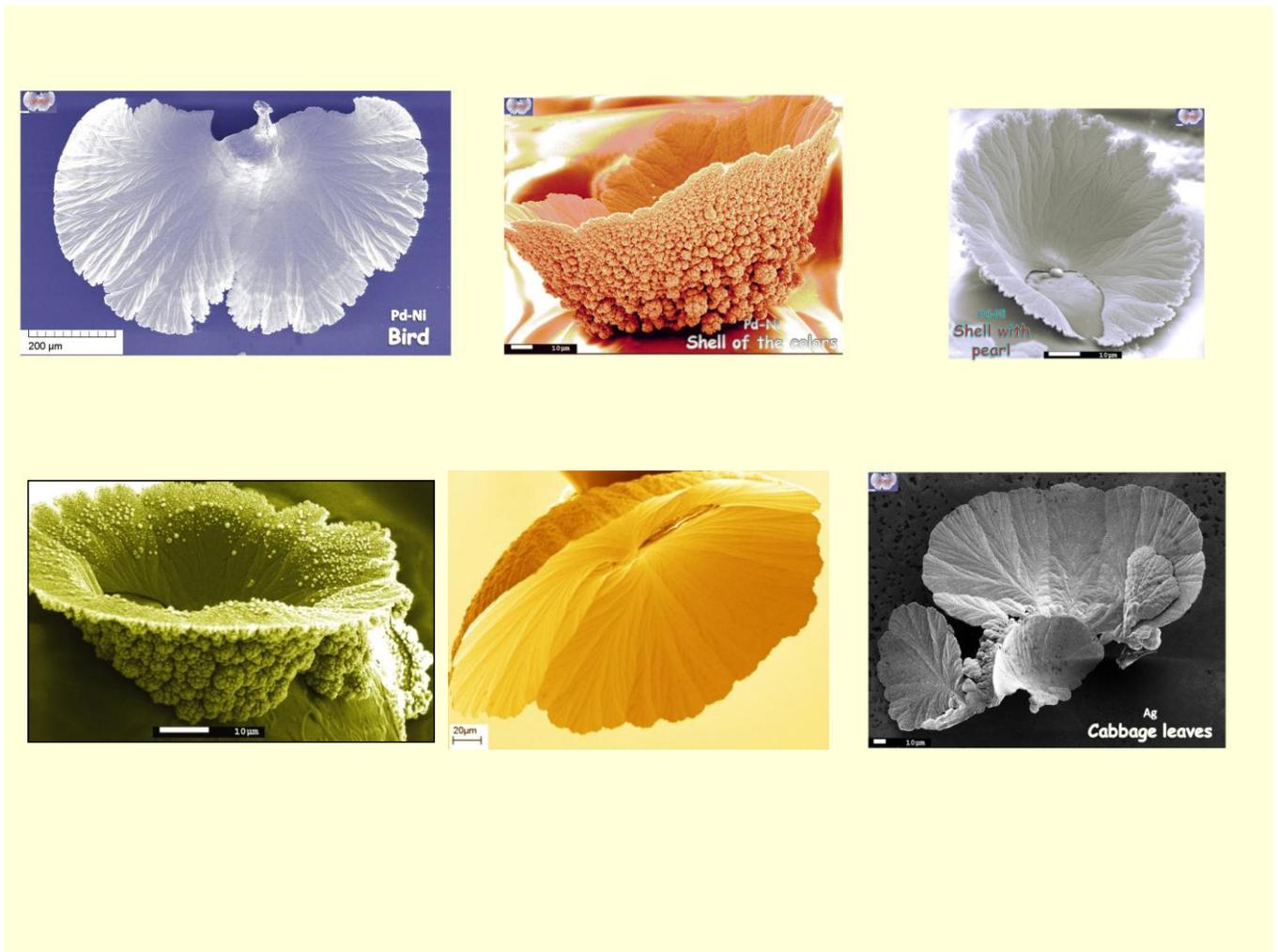

Fig. 2. Metallic "shells": a - "bird", b - "bowl", c -" flower", d - "hat", e -" shell with a pearl", f - "cabbage leaves".

Similar structures were obtained from copper, nickel, palladium and rhodium. The grown structures show several characteristic features. "Shell" walls start growing from the "root" ("bottom") which is a stub (or empty stub)-like site of layered fibers. The edge of the site is a circle or a nonclosed horseshoe-shaped line. The ends of the nanowire bundles often rise to the outer "shell" surface as "nanoflowers" (Fig.3) which can serve as templates for growing nanowires.

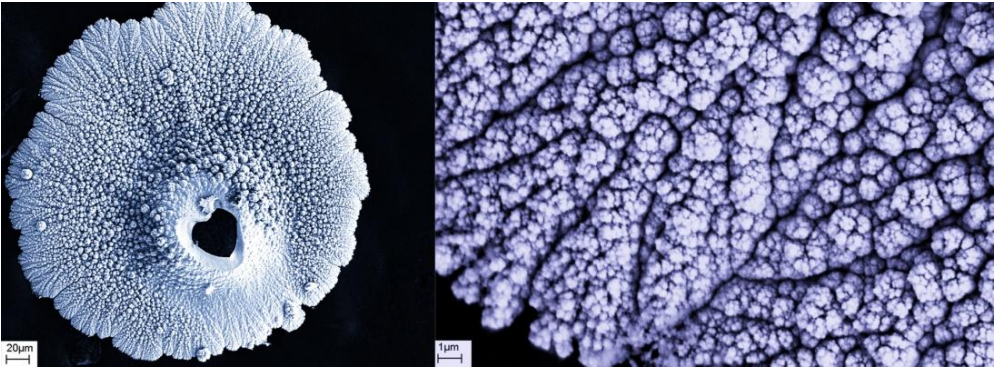

Fig.3. Outer "shell" surface with "nanoflowers".

Obviously the "shell" walls are formed by the conical nanowire bundles growing from this line, the bundles expanding upwards owing to branching. Such "brooms" growing from different sites were observed among the diverse bulk Pd-Ni structures on membranes with disordered pore (Fig.4a). Similar "brooms" were isolated upon fragmentation of the "shell" in the ultrasonic bath (Fig.4b).

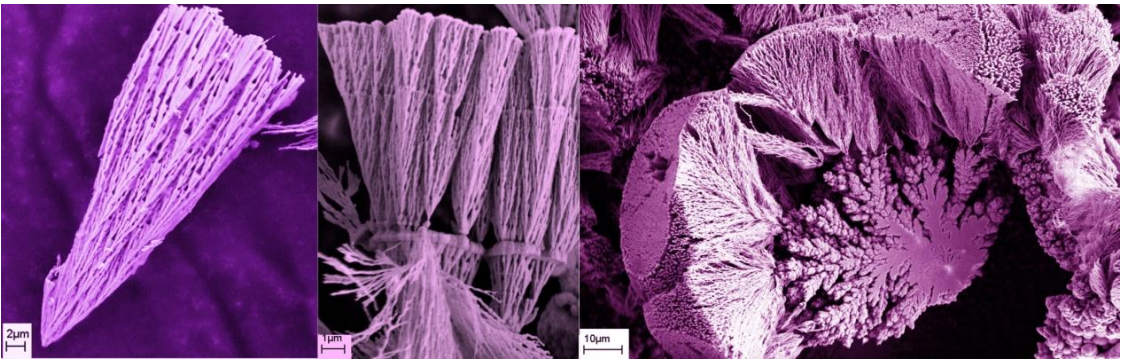

Fig.4. a - "broom", b - upon fragmentation, c - "broom" growth from the "shell" center.

The inner surface of the shell has a roughness at the nano- level and relief at the micro-level. Regular roughness at the nano- level becomes visible via the immersion gold plating (Fig. 5). The inner "lotus leaf" and "shell" surfaces exhibit a nonuniform relief, a pronounced woven pattern (Fig.6), its lines directed from the "root" to the periphery, as well as transverse rings (Fig.7).

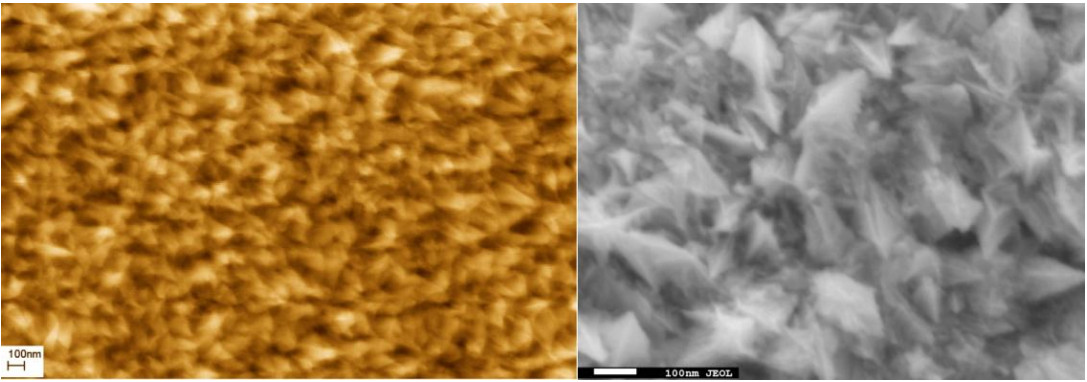

Fig.5. Inner Pd-Ni "shell" surface with immersion gold coating.
The nanostructure shows the regular surface roughness.

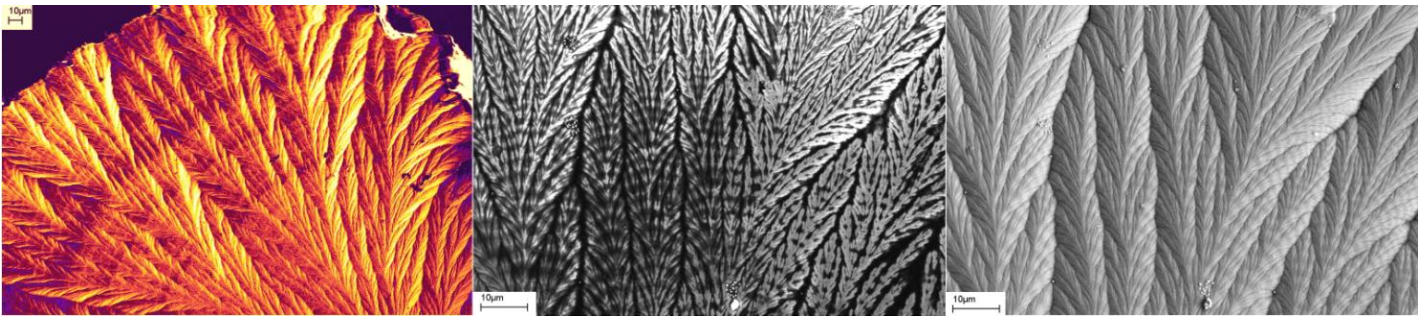

Fig.6. a – woven pattern on the inner "shell" surface; b - was obtained using a high resolution in-lens secondary electrons detector.

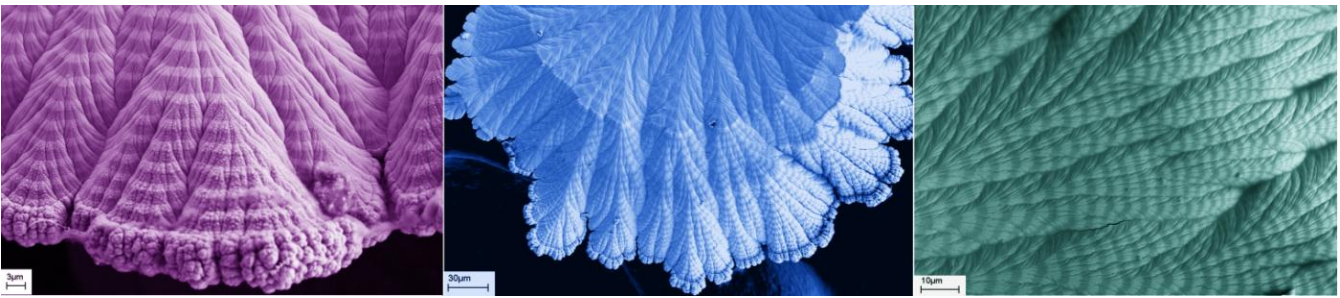

Fig.7. Transverse rings on "shell" surfaces.

The grown "leaves" and "shells" are examples of metallic woven multilayer surfaces. This relief is a manifestation of the inner "shell" architecture on the surface layer. Such architecture is observed if the surface layer has been removed mechanically (Fig.8) or by means of chemical etching (Fig.9).

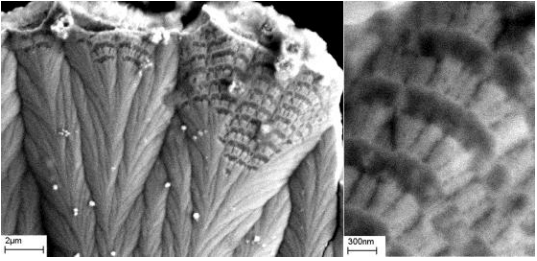

Fig.8. Inner "shell" surface. The disturbed surface layer shows "bricks", conical elements with circular bands ("bandages").

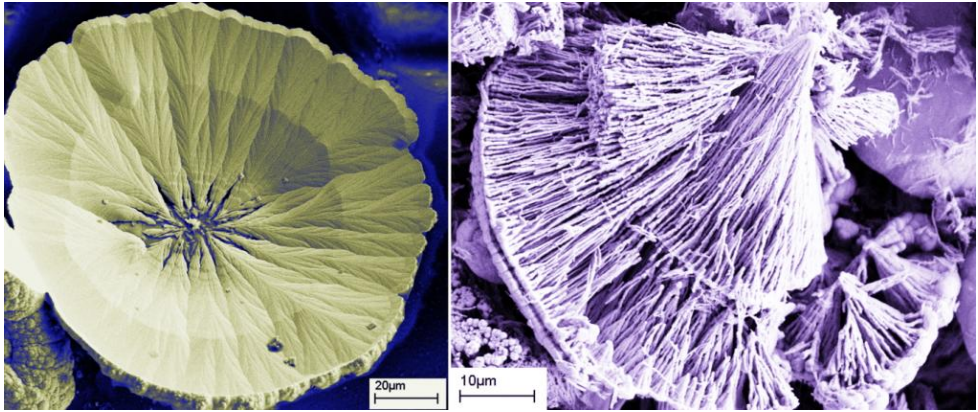

Fig.9. Shell" before and after fragmentation with simultaneous etching.

The images of the pieces obtained by fragmentation of the metallic "shell" in the ultrasonic bath (Fig.10) and chemical etching (Figures.11-13) exhibit a pronounced fractal hierarchic structure.

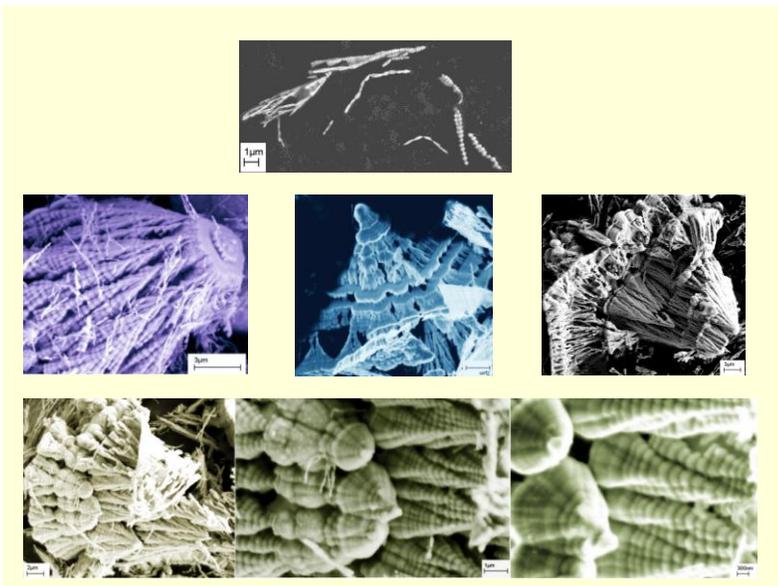

Fig.10."Shell" pieces upon fragmentation in ultrasound bath. One can see the hierarchic layer structure: submicron conical elements, "bricks", and the conical microstructures grown upon their fractal branching.

It is seen that the basic building blocks of metallic "shell" are nanowires with diameters of tens nanometers (Fig. 10, 13). These wires have the periodic thickenings like the beads.

The metallic "shell" have a bowl-shaped frame, its walls composed of layers of densely packed nanoelements. Each nanoelement is a conical bundle of nanowires grown as a supported "wine glass".

The shape of the "shell" frame (Fig.2, "bowl") copies the shape of the fragments (Fig.10, "jelly-fish") and the nano- building elements, the conical nanowire bundles (Fig.11), i.e., the hierarchic structure is reproduced at the nano-, micro- and meso-levels.

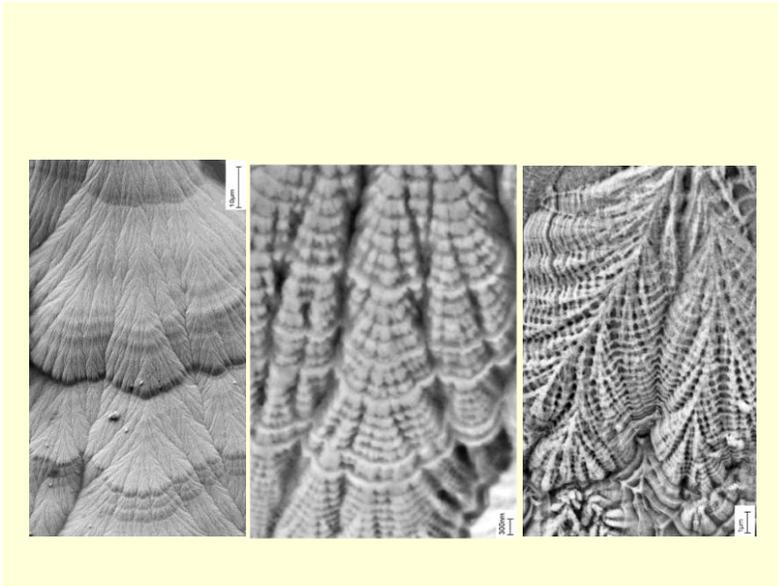

Fig.11. a - Inner Pd-Ni "shell" surface, b - surface after 10 s chemical etching, c - after 30 s chemical etching. The hierarchic structure is seen at the nano- and micro-levels.

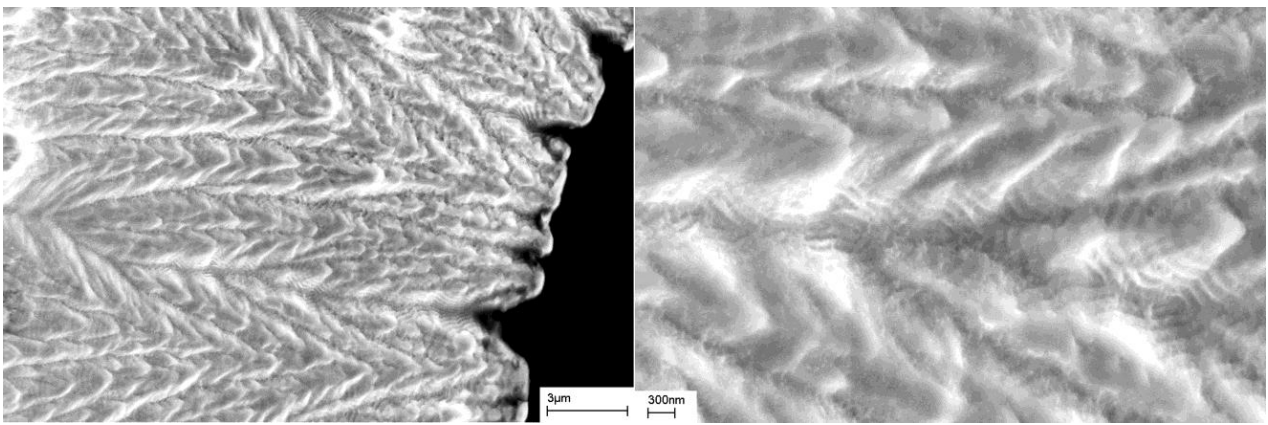

Fig.12. Inner "shell" surface upon chemical etching. The hierarchic layer structure is seen at the nano- and micro-levels.

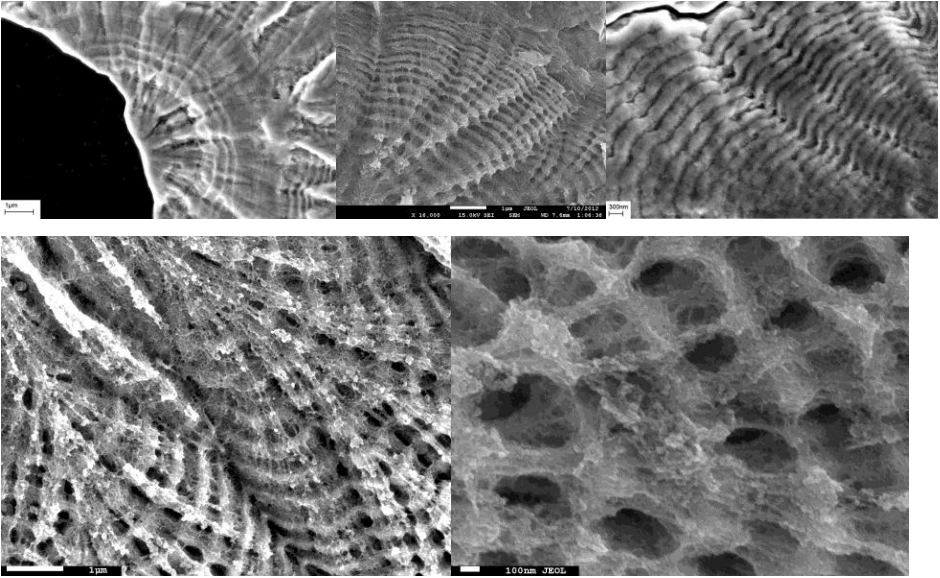

Fig.13. Etching pattern of the inner "shell" surface: fractal networks with circular bands ("bandages").

Numerous investigations have shown that the excellent mechanical properties of natural materials are due to their complex architecture and hierarchic structure at the nano-, micro- and meso-levels [7]. The natural nacre hierarchical scheme is called "bricks and building mortar". The superb strength, viscosity and resistance of this material are achieved by the reliable nanostructure of alternating layers of natural organic polymer, protein (10 -50 nm), and aragonite pellets (200-900 nm) 5-8 mkm in diameter. The hierarchic structure of our metallic "shells" is more complex owing to the more complex structure of the elementary "bricks" and the resulting three-dimensional lattice. Obviously, for more detailed study of the nano-structure is required the high resolution electron microscopy technique. The first TEM data revealed the essential distinction of the crystal state and phase composition of the "bricks" and the layer between them. The results of the research (to be published later) suggest that the use of various modes of pulse current and electrodeposition of various metals from two baths enables to implement the biomimetic approach in creating a model of natural nacre. We have managed to grow "shells" from Pd-Co alloy and copper which extends the set of feasible materials. Yet, the question is still open: what model sizes can be grown by the proposed method so that the hierarchic structure is preserved at the nano-, micro- and meso-levels. So far we have grown a 1.2 mm "shell" from Pd-Ni alloy.

### 3.2. Possible mechanisms of metallic model formation

The method of producing nanowires by dc current electrodeposition of metal in a porous membrane at controlled potential has long been known and extensively used. As a rule, the process of nanowire growth is finished immediately upon nanowire rise to the membrane surface following which the membrane is dissolved and the nanowires are isolated for targeted use. If the process is continued after nanowire rise, the membrane surface is healed over and used as a contact pad for measuring the wire resistivity. In those cases the nanowires produced no mesastructures.

We have grown nanostructured "vegetable" metallic meso-samples by pulse current electrodeposition continued after the nanowires appear on the membrane surface. The function of the pulse current mode is a portionwise addition of metal, i.e., by metal clusters. The crucial feature of pulse current electrodeposition is the possibility to carry out the process at higher current density than during dc current deposition and to achieve metal deposition in the form of nanocrystals. The latter enables to produce hierarchic structures at the nano- level.

The template is the other essential component of our method.

The nanowires growing from the membrane serve as the first template during mesostructure growth. The surface with a deposited metallic pattern or regular etching pits can also act as a template. As a result of self-assembly, metal nanoclusters are deposited on the templates form the elementary "bricks". The images obtained (for instance, "jelly-fish", etc.) demonstrate that the architecture of the metallic "shell" is layered, and each subsequent layer grows independently using the roughness of the previously built layer as templates. Pulse current electrodeposition on templates is obviously a tool of forming a hierarchic structure of grown metallic models.

The typical works on synthesis of "nanoflowers" (by chemical deposition [11, 13, 14,15] as well as electrodeposition [10])  have one feature in common: material is deposited on an as-prepared porous membrane [10], or deposition is preceded by formation of a porous polymer template (matrix) [15], a cellular metal frame [13] or a zinc oxide sublayer with specific morphology [11]. These facts suggest that the known examples of nanoflowers" are particular cases of the general method of forming bio-inspired models during irregular (pulsed) growth of material on porous membranes or templates. For instance in [15], the authors obtained a silver "leaf" on a polymer matrix by chemical reduction of silver solution which does not differ in shape from the structure presented in [8] grown on an aluminum oxide membrane by electrodeposition from an other electrolyte at a specified pulse current mode.  It is clear that in the above cases the irregular, pulsed deposition was achieved spontaneously due to the concentration inhomogeneities in the reaction system. The authors did not study the effect of the deposition modes and template shapes on the model structure which did not allow determining the general pattern of the method among the numerous separate cases and controlling the model shape.

By the physics of the process, its technical execution and the final result the method proposed is principally different from the known "electrospun" method [1]. It is suggested that models are grown from metals and alloys by electrodeposition from electrolyte solutions which makes it possible, in particular, to produce composite coatings by electrodeposition from dispersed phase suspension (for instance, diverse ceramics) in metallization electrolyte. The process is carried out in an electroplating unit on membranes (templates) using pulse current by a specified program. The one-stage process results in mesoscopic models of biological objects with a hierarchic structure at the nano-, micro- and meso-levels. In the "electrospun" method melt or polymer solution extrusion through a die in electric field gives rise to "mops" consisting of chaotically tangled polymer fibers. It also allows obtaining hybrid material by PVD deposition of copper on polymer fiber [16]. Silver tubes and micron-sized copper and nickel items are produced from polymer fibers by a sequential scheme including precursor treatment, chemical reduction of metal followed by removal of polymer and annealing. In contrast to the technique proposed, the "electrospun" method as it stands is not essentially biomimetic, i.e. it does not model natural objects or processes.

### 3.3. Morphogenesis Hypothesis

The shape resemblance of the synthesized metallic mesostructures to biological objects is flaring obvious which is illustrated in Fig.14 by several examples.

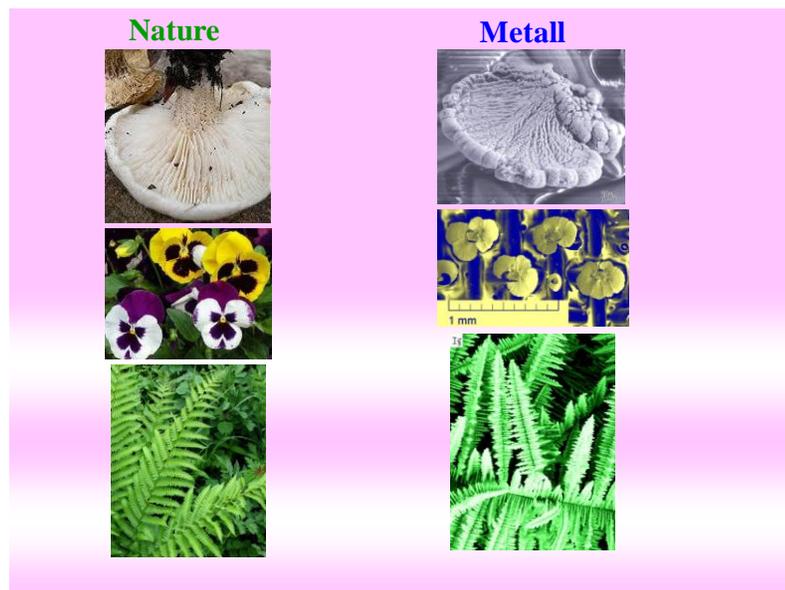

Fig 14. Natural objects and metallic models.

As shown above, the synthesized metallic models also exhibit a hierarchic structure inherent in natural objects. We believe that the shape resemblance of our grown metallic mesostructures to biological objects is not accidental, it is likely to be specified by the coincidence of some essential mechanisms of the growth processes and shaping. We can definitely identify several important general characteristics of the growth processes and shaping of fungi and plants and their metallic models.

1) Growth occurs on membranes or templates.

2) Pulsed or irregular growth.

3) Growth is achieved by addition of material in certain growing points, the so-called "growing tops" for

plants and "nuclei" in case of electrocrystallization. The location of the growing points is shape-determining and leads to fractal branching. It should be noted that practically all the grown metallic mesostructures are fractals. They are classical Mandelbrot fractals, "cauliflower" and "broccoli". "Fern" and 'shells" are also well expressed fractals. This aspect of growing metallic models by pulse current electrodeposition on porous membranes makes it also similar to biological morphogenesis.

The general characteristics of shaping bio-inspired objects and their metallic models as well as the diversity of the synthesized models and their external and structural resemblance to the biological prototypes reveal the general mechanism of morphogenesis [10].

We suppose that pulsed growth on membranes is a tool of morphogenesis for most mushrooms and plants which is accompanied by fractal branching and self-assembly of growing clusters and fibers. The species diversity is likely to be due to the diversity of biological membranes that are known to be synthesized in the course of self-assembly involving genetic codes and various pulsed growth modes. The latter, in turn, may depend on numerous factors such as solar intensity, water amount and chemical growth-promoting agents, etc.

## 4. Conclusion

The possibility of shaping metallic mesostructures during their pulse current electrodeposition on membranes has been shown: the conditions of synthesis of mushrooms- and shell-like convex-concave models have been found. The SEM research has revealed that the multilayer frame of the metallic "shell" has a hierarchic fractal structure made by self-assembly of conical nanowire bundles. Therefore, the method proposed enables to reproduce not only the shapes of natural objects (protozoa, plants, mushrooms, shells), but also their hierarchic structure at the nano-, micro- and meso-levels. This suggests that it is a biomimetic technique. The general characteristics of shaping bio-inspired objects and their metallic models as well as the diversity of the synthesized models and their external and structural resemblance to the biological prototypes suggest that pulsed growth on membranes is a tool of morphogenesis for most mushrooms and plants. The results obtained are certainly of interest for development of nano-scale self-assembly processes as well as for simulation of biological morphogenesis. Due to the technological simplicity, the grown nanostructured woven metal and alloy materials offer considerable promise for nanoplasmonics, fabrication of efficient catalysts, superhydrophobic surfaces for technical applications, medical filters, sensors.

## 5. Acknowledgments

The authors thank colleagues: Andrey Mazilkin , Ivan Veshchunov and Evgeniya Postnova for their great

help in this work.

6. **References**

**1.** Xianfeng Wang, Bin Ding, Jianyong Yu, Moran Wang. Engineering biomimetic superhydrophobic surfaces of electrospun nanomaterials (Review).

Nano Today, Vol. 6, Issue 5, 510-530, 2011.

**2.** Qian Li, Haikuo Sun, Ming Luo, Wenjian Weng, Kui Cheng, Chenlu Song, Piyi Du, Ge Shen and Gaorong Ha. Room temperature synthesis of ZnO nanoflowers on Si substrate via seed-layer assisted solution route. Journal of Alloys and Compounds, Vol. 503, Issue 2, 514-518 (2010).

**3.** Dan Xu, Jiangjiang Gu, Weina Wang, Xuehai Yu, Kai Xi and Xudong Jia. Development of chitosan-coated gold nanoflowers as SERS-active probes.

Nanotechnology, **21,** Issue 37, p.5101 (2010).

**4.** Xiaomei Chen, Bingyuan Su, Genghuang Wu, Chaoyong James Yang, Zhixia Zhuang, Xiaoru Wang and Xi Chen. Platinum nanoflowers supported on graphene oxide nanosheets: their green synthesis, growth mechanism, and advanced electrocatalytic properties for methanol oxidation.

J. Mater. Chem., **22**, 11284-11289 (2012).

**5.** Hongmei Zhang, Weiqiang Zhou, Yukou Du, Ping Yang and Chuanyi Wang,

One-step electrodeposition of platinum nanoflowers and their high efficient catalytic activity for methanol electro-oxidation. Solid State Sciences, Volume 12, Issue 8, 1364-1367, 2010.

**6.** Barthelat F. and Zhu D. J. A novel biomimetic material duplicating the structure and mechanics of natural nacre. J. Mater. Res. **26,** 1203 (2011).

**7.** I. Corni, T.J. Harvey, J.A. Wharton, K.R. Stokes, F.C. Walsh and R.J.K. Wood.

A review of experimental methods to produce a nacre-like structure (topical review).

Bioinspiration & biomimetics, **7** , pp 031001-24 (2012)

**8.** G.K. Strukova, G.V. Strukov, E.Yu. Postnova, A. Yu. Rusanov, A.D. Shoo.

"With no Color and Scent: Nanoflowers of Metals and Alloys" **arXiv:1108.4589**

**9.** G. K. Strukova, G. V. Strukov, A. Yu. Rusanov, S. V. Egorov. "With no Color and Scent (part II): Metal and Alloy Microstructures-Handmade Replicas of Natural Objects". **arXiv:1112.18**

**10.** G. K. Strukova, G. V. Strukov, S. V. Egorov. International Conference of Surfaces, Coatings and Nanostructured Materials, 27-30 March 2012, Tampa, Florida, USA. (Oral report).

**11.**   Huanhuan Kou, Xin Zhang, Yongling Du, Weichun Ye, Shaoxiong Lin and Chunming Wang. Electrochemical synthesis of ZnO nanoflowers and nanosheets on porous Si as photoelectric materials . Applied Surface Science, Volume 257, Issue 10, p. 4643-4649 (2011).

**12.**   Qian Li, Haikuo Sun, Ming Luo, Wenjian Weng, Kui Cheng, Chenlu Song, Piyi Du, Ge Shen and Gaorong Ha.  Room temperature synthesis of ZnO nanoflowers on Si substrate via seed-layer assisted solution route.  Journal of Alloys and Compounds, Volume 503, Issue 2, p. 514-518 (2010).

**13**. Linmei Li and Jian Weng.  <u>Enzymatic synthesis of gold nanoflowers with trypsin</u>.   Nanotechnology, **21** , Issue 30, p. 5603 (2010).

**14**.  Dan Xu, Jiangjiang Gu, Weina Wang, Xuehai Yu, Kai Xi and Xudong Jia.  Development of chitosan-coated gold nanoflowers as SERS-active probes.  Nanotechnology, **21,** Issue 37, p.5101 (2010).

**15**. Hongmei Zhang, Weiqiang Zhou, Yukou Du, Ping Yang and Chuanyi Wang. One-step electrodeposition of platinum nanoflowers and their high efficient catalytic activity for methanol electro-oxidation.  Electrochemistry Communications, Volume 12, Issue 7,  p.882-85 (2010).

**16.** Kai Wei, Tomohiro Ohta, Byoung-Suhk Kim, Kwan-Woo Kim, Keun-Hyung Lee, Myung Seob Khil, Hak-Yong Kim, Ick-Soo Kim. Development of electrospun metallic hybrid nanofibers via metallization.  Polymers for Advanced Technologies, Vol. 21, Issue 10, p. 746–751 (2010).

.